\documentclass[runningheads]{svmult}

\usepackage{makeidx}   
\usepackage{graphicx}  
\usepackage{subeqnar}  
\usepackage{multicol}  
\usepackage{physprbb}  
\makeindex             

\begin{document}
\title*{The Globular Cluster Luminosity Function: New Progress in
Understanding an Old Distance Indicator}
\toctitle{Globular Cluster Luminosity Function}
%
%
\titlerunning{Globular Clusters}
%
\author{Tom Richtler}
\authorrunning{Tom Richtler}
%
%
\institute{Astronomy Group, Departamento de F\'{\i}sica, Universidad de
Concepci\'on, Casilla 160-C, Concepci\'on, Chile}

\maketitle              

\begin{abstract}
I review the Globular Cluster Luminosity Function (GCLF) with emphasis on
recent observational data and theoretical progress. As is well known, the turn-over magnitude (TOM)
is a good distance indicator for early-type galaxies within the limits set by data quality and
sufficient number of objects. A comparison with distances derived from 
surface brightness fluctuations with the available TOMs in the V-band reveals,
however, many discrepant cases. These cases often violate the condition that the
TOM should only be used as a distance indicator in old
globular cluster systems.  
The existence of intermediate age-populations in early-type galaxies likely is the
cause of many of these discrepancies.
The connection between the luminosity functions of young and old cluster systems
is discussed on the basis of modelling the dynamical evolution of cluster systems.
Finally, I briefly present the current ideas of why such a universal structure as
the GCLF exists.

\end{abstract}

\section{Introduction: What is the Globular Cluster Luminosity Function?}

Since the era of Shapley, who first explored the size of the
Galaxy, the distances to globular clusters often set landmarks in
establishing first the galactic, then the extragalactic distance scale.
Among the methods which have been developed to determine the
distances of early-type galaxies, the usage of globular clusters
 is one of the oldest, if not {\it the} oldest. 
Baum \cite{baum55} first compared the brightness of
the brightest globular clusters in M87 to those of M31. 
With the observational technology improving it became possible to reach
fainter globular clusters and soon the conjecture was raised that the distribution of absolute magnitudes of globular
clusters in a globular cluster system exhibits a universal shape, which can
be well approximated by a Gaussian: 

$$ \frac{dN}{dm} \sim exp{\frac{-(m-m_0)^2}{2 \sigma^2}}, $$

where dN is the number of globular clusters in an apparent  magnitude bin $dm$, $m_0$ is
the ''Turn-Over Magnitude'' and $\sigma$ the width of the Gaussian distribution.
Also a representation by a ''t5-function''
$$ \frac{dN}{dm} \sim \frac{1}{\sigma} (1+\frac{(m-m_0)^2}{5 \sigma^2})^{-3} $$
introduced by Secker  \cite{secker92} found a wide-spread application.

This distribution is called the ''Globular Cluster Luminosity Function''. 
In the following ''Globular Cluster Luminosity Function`` is abbreviated by
GCLF, ''Turn-Over Magnitude`` by TOM, and ''Globular Cluster System`` by GCS.

The conjecture that the GCLF is very similar in different galaxies, in particular that the absolute magnitude 
of the TOM has an almost universal value, 
 has been first suggested  by Hanes \cite{hanes77} (but also see the references in this paper). The reviews of Harris \& Racine  \cite{hara79}
,
Hanes  \cite{hanes79}, Harris  \cite{harris91}, Jacoby et al.  \cite{jacoby92}, Ashman \& Zepf  \cite{ashman98}, Whitmore \cite{whitmore97b}, Tammann \& Sandage
\cite{tammann99},
 and Harris  \cite{harris01} demonstrated both the solidity and the limitations
of this conjecture.  They  also show the progress which has been achieved during
 the
past 20 years both in terms of the number of investigated GCSs  and the
accuracy of an absolute calibration. 

The GCLF as a distance indicator has seen little application to  spiral galaxies
for several reasons:
their GCSs are distinctly poorer than those of giant ellipticals, the identifications of clusters is rendered more difficult by the projection onto the disk,
and the presence of dust causes inhomogeneous extinction.  
The investigation of GCSs of ellipticals or S0-galaxies is much easier due to the homogeneous light background, the richness, and the absence of
internal extinction. 
 
The application of GCLFs as  distance indicators for early-type galaxies
has a  simple recipe: Given  appropriately deep photometry
of the host galaxy, identify GC candidates, as many as you can.
In most cases this has to be done statistically by considering only objects
with GC-like colors and by subtracting a hopefully well determined background
of sources. Then
measure their apparent magnitudes, draw a histogram 
 and fit a suitable function, for instance a Gaussian,
to determine the apparent TOM. Under the assumption  that every GCS
has the same absolute TOM, one can use the galactic system and/or the
M31 system to calibrate it in terms of absolute magnitudes. Real data, however,
make the derivation of the TOM somewhat more difficult, which will be discussed
below. The most important restriction is probably that we can follow the GCLF
down to faint clusters only in two galaxies, the Milky Way and Andromeda.

There is no physical reason why the GCLF should be a Gaussian or a t5-function.  On the contrary, we shall later 
on learn about physical reasons why it is  
{\it not} a Gaussian.
Closer scrutiny
of the galactic cluster system indeed  shows that its GCLF is not  
symmetric in that it exhibits an extended wing beyond the TOM for smaller
masses (for example see Fig.2 of Fall \& Zhang  \cite{fall01}). But Gaussians empirically are fair descriptions for the bright side and  
  most observations of GCLFs of distant
galaxies seldomly reach more than 1 mag beyond their TOMs, so this asymmetry
is not relevant for measuring the TOM. 
  
\begin{figure}[t]
\begin{center}
\includegraphics[width=.6\textwidth]{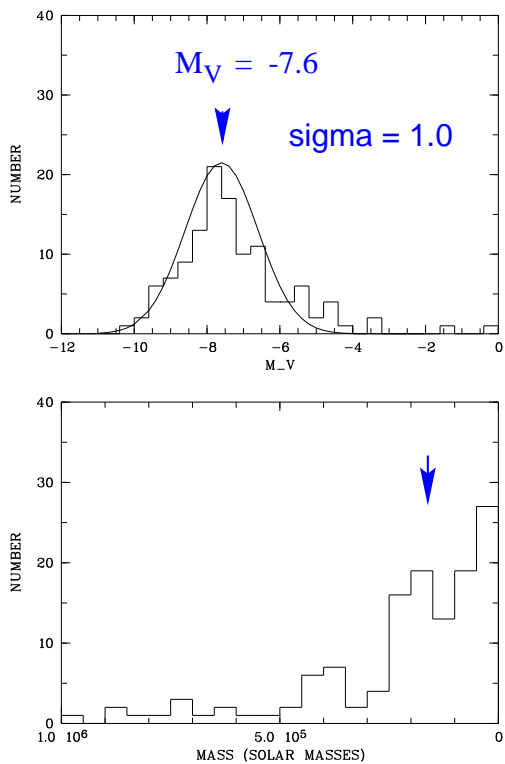}
\end{center}
\caption[]{The upper panel shows the globular cluster luminosity function
for the galactic system together with a fitted Gaussian to the distribution.
 Only clusters with reddening E(B-V) less than 0.8 mag,
and absolute magnitudes brighter than -4 have been considered in the fit. 
The lower
panel shows the mass distribution in linear mass bins. The mass corresponding
to the TOM is indicated.
} 
\label{milky}
\end{figure}

Fig. \ref{milky} shows the GCLF for the galactic system (upper panel). It has been constructed on the basis of the ''McMaster-catalog'' (Harris \cite{harris96}) using the horizontal
branch (HB) brightness as the distance indicator and adopting the relation
 $M_V(HB) = 0.2 \cdot [Fe/H] + 0.89$ (Demarque et al. \cite{demarque00}). The
Gaussian fit results in $M_V = -7.56 \pm 0.12$ for the TOM and $1.2 \pm 0.1$
for the dispersion of the Gaussian. The lower panel shows a histogram of the 
masses, assuming M/L = 2, where the mass which corresponds to the TOM is
indicated. In this linearly binned histogram, there is no striking feature 
at this mass. {\it Indeed, the existence of a TOM is a consequence of the logarithmic magnitude scale in combination with a change of the power-law slope of the mass function.}
We come back to this in a later section.

Throughout this review, we shall consider only TOMs in the V-band,
because most modern published data, particularly those from the Hubble Space
Telescope, have been obtained in V. Other photometric systems, most notably the Washington
photometric system (Geisler et al. \cite{geisler96}, Ostrov et al. \cite{ostrov98}, Dirsch et al. \cite{dirsch03}) have been used for the investigations of GCSs
as well.

Given the previous excellent reviews on GCLFs as distance indicators, what
can be the scope of this contribution? A lot of new data has been published
during the last years and it is now possible to compare GCLFs of early-type
galaxies with other distance indicators on the basis of a much larger sample
than has been possible before.
The outstanding publication  here is the catalog of distances based on
surface brightness fluctuations (SBFs) (Tonry \cite{tonry01}).  We shall see that    
the absolute calibration of GCLFs indeed agrees very well with that of
SBFs, demonstrating that  {\it most} GCSs of elliptical or S0-galaxies 
show absolute TOMs which are not distinguishable within the uncertainties of the measurements. 
Nevertheless, many discrepancies between GCLF distances and SBF distances
exist. These galaxies are of particular interest and we shall discuss them
as well. 
 
Beyond the usefulness of the GCLF as a distance indicator is the question
why there exists such a remarkably universal structure. Is there a universal
formation law for globular clusters, which operates in the same way in such
different galaxies as the Milky Way and giant ellipticals? 
This problem has to do with  the initial mass function of globular clusters and
the evolution of GCSs. Much progress has been achieved during the last
years, on which we will also report.

\section{Sources of uncertainty}

\begin{figure}
\begin{center}
\includegraphics[width=.9\textwidth]{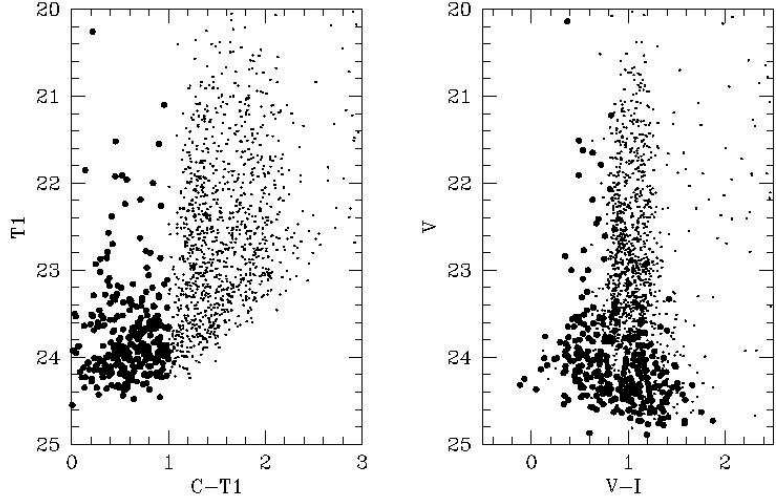}
\end{center}
\caption[]{This plot has been taken from Dirsch et al. \cite{dirsch03}. It
shows a comparison of the colour-magnitude diagrams of the GCS of NGC 1399
in the Washington system and in V-I. The many unresolved background galaxies
(a few stars are mixed in as well) with an excess in the blue filter C fall outside the colour-range of the globular
clusters in C-T1, whereas they populate that range in V-I and so add significantly to the background at faint magnitudes.}
\label{wash}
\end{figure}

To begin with difficulties: Even if we trust the universal TOM, the
actual measurement may appear straightforward according to the above recipe,
but nevertheless one encounters many sources of uncertainty. The identification
of GCs as resolved objects from the ground is only possible for the nearest
early-type  galaxies, for
example NGC 5128 (Rejkuba \cite{rejkuba01}). Unfortunately, no modern
investigation of the GCLF of  NGC 5128 exists until now. Observations with the Hubble Space Telescope can resolve the largest clusters in  galaxies as distant
as about 20 Mpc (e.g. Kundu \& Whitmore
 \cite{kundu01},  \cite{kundu01b},  Larsen et al.  \cite{larsen01}).
Therefore, the identification by ground-based
observations normally has to use
color criteria and the statistical discrimination against a ''background'', which actually
may consist of foreground stars and of unresolved background galaxies.
That the latter contamination is a strong function of the color system used,
is nicely shown in Fig. \ref{wash}, taken from Dirsch et al. \cite{dirsch03}, which compares the color magnitude diagrams
 V-I and Washington C-T1
for the GCS of NGC 1399, the central galaxy in the Fornax cluster. The
C-T1 color recognizes many background galaxies, which have an excess flux
in the blue band C, while they are not noticable in V-I, a color which has
been widely used in HST investigations.

\begin{figure}
\begin{center}
\includegraphics[width=.5\textwidth,angle=-90]{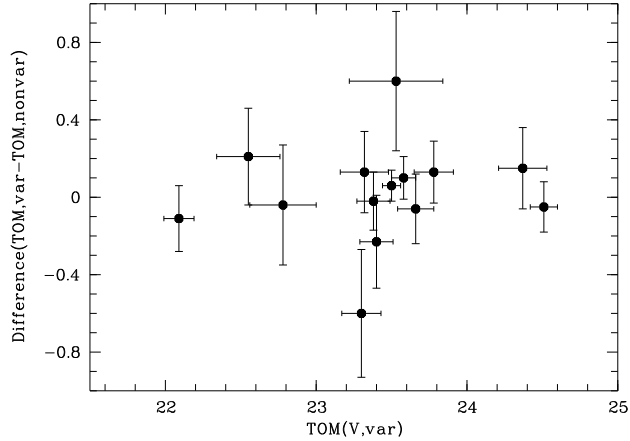}
\caption[]{This plot is based on HST observations of a sample of 14 early-type
galaxies by Larsen et al. \cite{larsen01}. It shows the differences of the TOMs derived from
the t5-function fits leaving the width either variable or non-variable versus the TOM derived
by fits with a variable width. The two most deviating galaxies are NGC 1023 and NGC 3384.   
}
\end{center}
\label{larsen}
\end{figure}

The nearest large galaxy clusters (Virgo and Fornax) have
  distance moduli of about 31. Thus, TOMs for GCSs  at this
  distance and beyond will generally be fainter than V=23.5,
 where the photometric incompleteness (depending on the data quality) plays an increasingly dominant role. Then there are the factors of 
the numbers of found GCs and the distance itself: If the photometry
does not reach the TOM and/or the GCS is not very rich, the resulting
TOM is naturally less well defined than in the case of a nearby, rich GCS. 
Then it may depend on the adopted shape of the luminosity function. 
However, 
empirically the derived TOM is not very sensitive to whether a t5-function
or a Gaussian is used, at least not in the case of well observed GCLFs (e.g.
see Della Valle et al. \cite{della98} for the GCLF of NGC 1380).

 Moreover, the
width of the adopted fitting function can be left free or can be fixed.  
Larsen et al. \cite{larsen01} performed t5-function fits to their sample of 14 early-type
galaxies both with the width as a fit parameter and with a fixed with of $\sigma_t = 1.1$ mag.
To gain an impression of the effect on the TOM, we show Fig.\ref{larsen}, where the TOM 
corresponding to a free width is plotted versus the difference TOM(var)-TOM(nonvar).   
Leaving the two outliers NGC 1023 and NGC 3384 aside, the standard deviation of the differences
is 0.13 mag. Then there are different ways to fit, for example maximum likelihood methods or direct fits.

One must also not forget the uncertainty of the foreground absorption and another factor which
is difficult to nail down: the photometric calibration of the respective
data set and the actual realization of the used photometric standard system.

The above error sources are always there, even if the TOM would be strictly universal, which one
would not expect: 
 For a given mass the luminosity of a GC depends on its
metallicity in the sense that metal-poor clusters are brighter in the
optical (e.g. Girardi
\cite{girardi02}), but the metallicity distribution within a cluster system is
not expected to be the same for all early-type galaxies. So one should 
principally correct for this as well (Ashman et al. \cite{ashman95}). There is also empirical evidence for
a metallicity dependence of the TOM: Larsen et al. \cite{larsen01} find by HST observations in V and I of a sample of 15 early-type galaxies the TOMs for red clusters to be fainter than those for
blue clusters by $\Delta m_V \approx 0.4$ mag, which is somewhat larger than predicted by theory.  
However, as we shall see, it is possible that part of this difference is due to an intrinsically
fainter TOM of the red cluster population, so it is difficult to quantify the metallicity effect.

Also if the initial cluster mass distribution would be the same in all galaxies, destruction processes like disk shocking or evaporation are expected to act
differently in different environments and may create intrinsically varying TOMs
(see section 10 for this topic).
Last, but not least, one has to assume that the members of a GCS all have
the same old  age, while we shall see that the number of examples 
 where this is not the case is growing. 

Given all these possible error sources, it  may come as a surprise that
GCLFs seem to work so well as distance indicators, and it is plausible
 that an accuracy of, say, 0.2 mag or less can only be achieved in the case of
rich GCSs and a high quality dataset. 

\section{The Galaxy and M31}
The two massive galaxies where the GCLF can be best observed down to faint clusters
are the Milky Way and M31. A calibration of the absolute TOM therefore via
these galaxies always
has to face the caveat that both are spiral galaxies and the application to
early-type galaxies may not be justified. However, as we will see, the zero-point gained from using the Galaxy and M31 as fundamental calibrators is in very
good agreement with the one obtained from the comparison with the method
of surface brightness fluctuations (Tonry et al. \cite{tonry01}), which at present offers the largest
and most homogeneous catalog of distances to early-type galaxies. 

The history of the investigations of the galactic GCS and that of M31 involves
the work of many people. To be short, we refer the reader to Harris \cite{harris01}  and
Barmby et al. \cite{barmby01} (and references therein) for the Galaxy and M31, respectively.
Harris \cite{harris01} quotes $M_V = -7.40 \pm 0.11$ for the galactic TOM and $\sigma =
1.15 \pm 0.11$. Barmby et al. \cite{barmby01} quote for the apparent TOM of the M31 system
$m_V = 16.84 \pm 0.11$ and $\sigma = 1.20 \pm 0.14$. The distance modulus of
M31 is $m-M_{M31} = 24.44 \pm 0.2$ (Freedman \& Madore \cite{freedman91}), which translates into an
absolute TOM of $M_V = -7.60\pm0.23$. The weighted average of these two TOMs
is $-7.46 \pm 0.18$, which we will compare with the distance moduli derived
from surface brightness fluctuations.

\section{The data}

During the last few years, many new TOMs of early-type galaxies
in the V-band have been published. The majority of them are based on HST observations and
stem from the papers by Kundu \& Whitmore \cite{kundu01}, Kundu \& Whitmore
\cite{kundu01b},
and Larsen et al. \cite{larsen01}. Other new papers on individual galaxies are from Okon \& Harris 
\cite{okon02},
 Kavelaars et al. \cite{kavelaars00}, Woodworth \& Harris \cite{woodworth00},
 Drenkhahn \& Richtler \cite{drenk99}.
 For TOMs published earlier we refer
the reader to the compilation of Ferrarese et al. \cite{ferra00} and references therein.
 As mentioned 
above, the main problem with such a data set is its inhomogeneity for a variety
of reasons. For example, Kundu \& Whitmore \cite{kundu01} and Kundu \& Whitmore
\cite{kundu01b} fitted Gaussians with both variable and fixed dispersions (1.3 mag) to their GCLFs. We adopt their TOMs resulting from the fixed dispersions because
of the larger number of galaxies included, leaving out a few TOMs with very
large uncertainties. 
Larsen et al. \cite{larsen01} fitted t5-functions with both non-variable and variable widths,
from which we adopt the latter because the scatter of the dispersions points to
real differences. However, the TOMs are not strongly influenced by whether the
dispersions are fitted or keep fixed. 
Since we are interested rather in the bulk properties of the available
data than in hand-selected data according to certain quality criteria, we
included also work which was mentioned but rejected by Ferrarese et al.

We end up with 102 TOMs (corrected for foreground extinction and including
a few double and triple measurements) in the V-band for 74 galaxies, which should be almost complete
from the present day back to 1994.

\section{The Hubble diagram}

\begin{figure}
\begin{center}
\includegraphics[width=.9\textwidth,angle=-90]{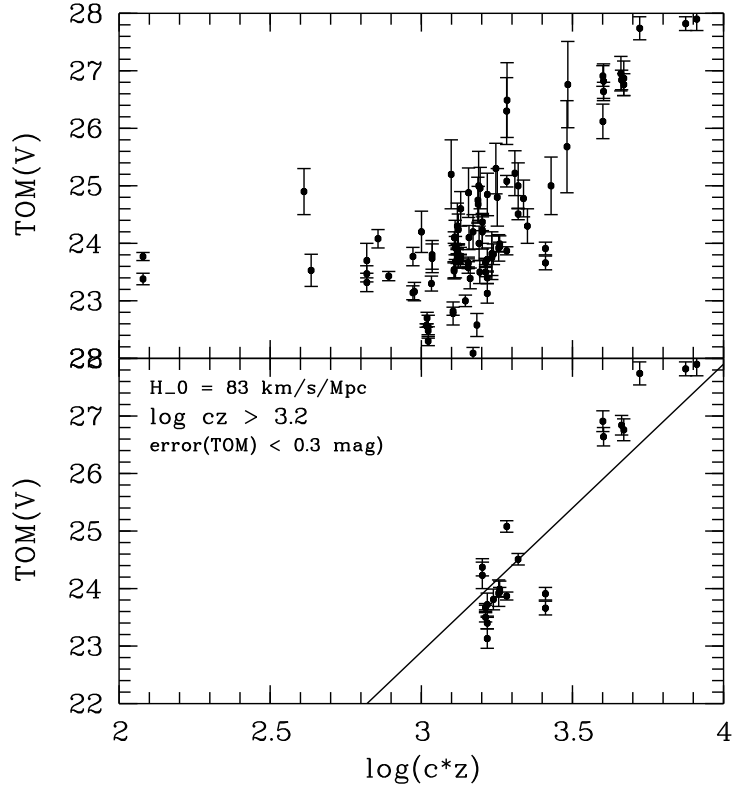}
\end{center}
\caption[]{The upper panel shows all of our collected TOMs versus their recession
velocities, defined here as the radial velocities related to the microwave 
background. Many of these galaxies obviously have peculiar velocities of
 the same order
as their recession velocities, in which case this definition is not adequate. The lower
panel selects those galaxies with TOM uncertainties less than 0.3 mag and log(cz) larger than 3.2 to reveal a Hubble constant of 83 km/s/Mpc. Also here, the TOMs do not define very well a slope of 5 in the  Hubble
diagram. Note that the group at V=27 are not directly measured TOMs, but deduced
from surface brightness fluctuations, see Lauer et al. \cite{lauer98} and Kavelaars et al. \cite{kavelaars00}.} 
\label{hubble}
\end{figure}

Can we say something about the Hubble constant from our data set, assuming that the
TOM indeed has the universal value of $M_V = -7.46 \pm 18$, adopted from the Milky Way
and M31?  
A set of standard candles whose redshifts are only due to their recession
velocities give a straight line in the Hubble diagram when their apparent
magnitudes (their TOMs in our case) are plotted versus their redshifts
according to
$$ m = 5 \cdot log(c \cdot z) - 5 \cdot log(H_0) + M - 25, $$
where $H_0$ is the Hubble constant in units of km/s/Mpc and M the constant 
absolute magnitude of the standard candles.

The upper panel of Fig.\ref{hubble} shows the Hubble diagram for our entire database. The velocities
of the host galaxies have been {\it individually} related to the
microwave background (which of course is not a good approach).  
It is obvious that
such a diagram is not suitable for deriving the Hubble constant. Many
objects show radial velocities which are simply not  in the
Hubble flow, most strikingly for NGC 4406 (which is represented by the double
measurement with the lowest velocity). If we select  galaxies with
log $c \cdot z > 3.2$ and furthermore only those with uncertainties  
less than 0.3 mag, we end up with about 20 galaxies. If these galaxies are
used to calculate the zero point in the Hubble relation, it gives
a Hubble constant of 83 km/s/Mpc, adopting a TOM of $\rm M_V = -7.5$ mag. It is clear that
 one cannot be content with this. Standard candles in a Hubble diagram should define
a straight line with a slope of 5, whereas the slope in this diagram is clearly steeper. To resolve this discrepancy, one has to carefully look into each individual GCS, select those TOMs with the highest degree of trustworthiness, and
then investigate the recession velocities of individual galaxies.
The measured radial velocities of galaxies within the space volume under
consideration may not be good indicators for their recession velocities due to the existence of large scale peculiar motions, which are under debate (e.g.
Tonry et al. \cite{tonry00}). 

Following Kavelaars et al. \cite{kavelaars00}, a better way might be to consider only groups of galaxies,
average the TOMs and assign a recession velocity to each group. Kavelaars et al. use
the Virgo, the Fornax and the Coma cluster and arrive at $69 \pm 9$ km/s/Mpc
for the Hubble constant. But to fix the recession velocities 
even for these three galaxy clusters is far from trivial. 

To avoid very lengthy discussions, a better way of deriving the Hubble
constant is perhaps the use of standard candles which are so distant that
peculiar velocities act only as minor perturbations of the Hubble flow, i.e.
the Hubble diagram of Supernovae Ia (Freedman et al. \cite{freedman01}).


\section{The comparison with surface brightness fluctuations}

\begin{figure}
\begin{center}
\includegraphics[width=.7\textwidth,angle=-00]{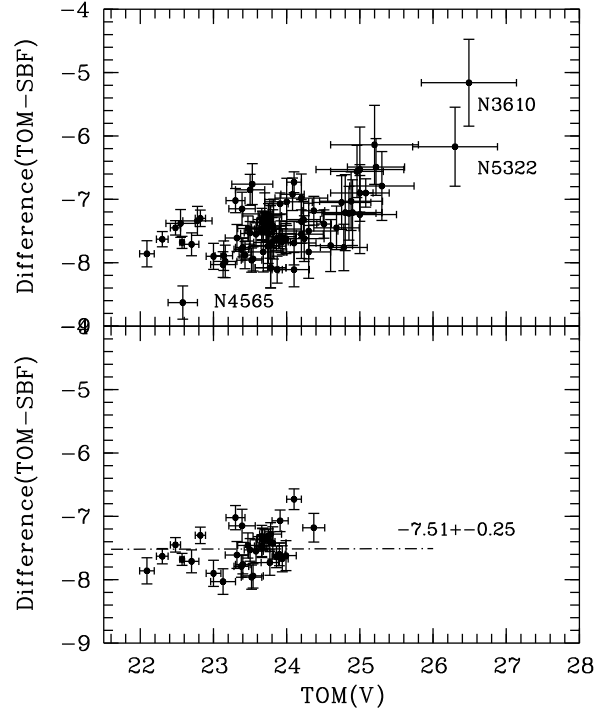}
\end{center}
\caption[]{The upper panel shows the difference between the TOMs and the
SBF distance moduli. For faint TOMs exists a trend to overestimate the distance
with respect to the SBF distance. The lower panel selects those galaxies with
uncertainties of both the TOM and the SBF distance less than 0.2 mag. The
mean difference agrees very well with the absolute TOMs from the Milky Way and
M31.
}
\label{SBFcomp}
\end{figure}

To evaluate the accuracy and reliability of the method of the GCLF, we
must compare it with other distance indicators of early-type galaxies.
This results in a complicated task, if one's objective is to select the
most reliable measurements, to quantify possible biases inherent to different
methods, and to discuss the uncertainties claimed by the authors. See for example Ferrarese et al. \cite{ferra00}, who conclude that GCLFs do not provide reliable
distances, mainly based on a deviating distance to the Fornax cluster, and 
Kundu \& Whitmore \cite{kundu01}, who contrarily find GCLF distances as least
as accurate as distances from surface brightness fluctuations. We do not want
to follow these lines but rather investigate what can be seen from the 
entirety of TOMs if they are compared with a distance indicator which 
provides distances to most of our GCSs.

Today, 
the most homogeneous and largest sample of distances to early-type galaxies
is the catalog resulting from the survey of surface brightness fluctuations
(SBFs) (Tonry et al. \cite{tonry01}; see also the preceding papers by Tonry et
 al. \cite{tonry97}, Blakeslee et al. \cite{blakeslee99}, and Tonry et al. \cite{tonry00}) which contains distances to about 300 galaxies.
Therefore  we restrict ourselves to a comparison with this important distance
indicator. Basically, it analyzes that part of the pixel-to-pixel scatter of a CCD image
of an early-type galaxy which is caused by the finite number of bright unresolved stars covered
by each CCD pixel. These fluctuations of the surface brightness are large for nearby galaxies and
small for more distant galaxies.

 Fig. \ref{SBFcomp} plots for all galaxies in our database (irrespective of
whether there are double or triple measurements) the TOM versus its
difference to the distance moduli from Tonry et al. \cite{tonry01}. The
error bars of the differences simply are the square roots of the quadratic
sums of the uncertainties in the GCLF and SBF distance moduli.
 The first impression
seems to be somewhat discouraging. Where we would have expected to see a
horizontal line at an ordinate value of -7.5 with some scatter, we see a
large spread with often dramatic devations, particular for the fainter TOMs.
What is striking is that the deviating galaxies do not scatter symmetrically
around a mean value, but that the faint TOMs give systematically larger
distance moduli than do the SBFs.
A direct and naive conclusion could be that perhaps  the very faint TOMs are
observationally not reached and that an extrapolation from the bright end
of the luminosity function to the TOM gives a TOM which is systematically
too faint. In fact this is not the case and the strongly deviating TOMs belong
to 
interesting galaxies (we come back to this point).

But also at the bright end there are irritations. The deviating galaxy at -8.6 is NGC
4565, and even the one with the brightest TOM, the Sombrero galaxy NGC 4594, does not fit very well
to our assumed universal value. Both are the only spiral 
galaxies in our sample. We note that the GCS of NGC 4565 
is very poorly populated  (Fleming et al. \cite{fleming95}), so this deviation might not bear much significance.

\section{Absolute TOMs and the distances to Virgo and Fornax}

However, if we select according to the quoted uncertainties, the situation
starts to look better. The lower panel of Fig.\ref{SBFcomp} plots all galaxies where
the uncertainties  of both the SBF distance and the TOM according to
the various authors are lower than 0.2 mag. The dispersion of the scatter is
0.25 mag and thus is compatible with the claimed selection. Thus we can confirm
the statement by Kundu \& Whitmore \cite{kundu01b} that the GCLF distances,
at least for the sample under consideration, are not less accurate than
the SBF distances. The mean  difference
is -7.51 mag with a dispersion of 0.24 mag and thus in excellent agreement with the zero-points coming
from the Milky Way and from M31. These three zero-points give a weighted mean
of $-7.48 \pm 0.11$. 

The average TOM of 8 galaxies in the Fornax cluster is 23.79 mag with a
dispersion of 0.17 mag, the one for the Virgo cluster (16 galaxies) is
23.62 with a dispersion of 0.16 mag, which translate into distance moduli
for Fornax and Virgo of $31.27 \pm 0.2$ and $31.10 \pm 0.2$, respectively.  
The corresponding distance moduli from the SBFs are $31.02 \pm 0.15$ and
$31.49 \pm 0.12$.
A discussion of the absolute calibration is not our objective. However,
we can 
  conclude that indeed many 
GCLFs can provide good distances but one is reluctant to label the GCLF ''universal'' at this point because there are too many deviations with the SBF distances.
 We shall see that these deviations are apparently related to the existence of
intermediate-age populations in early-type galaxies.

\section{Deviations and intermediate-age populations}

\begin{figure}
\begin{center}
\includegraphics[width=.7\textwidth,angle=-00]{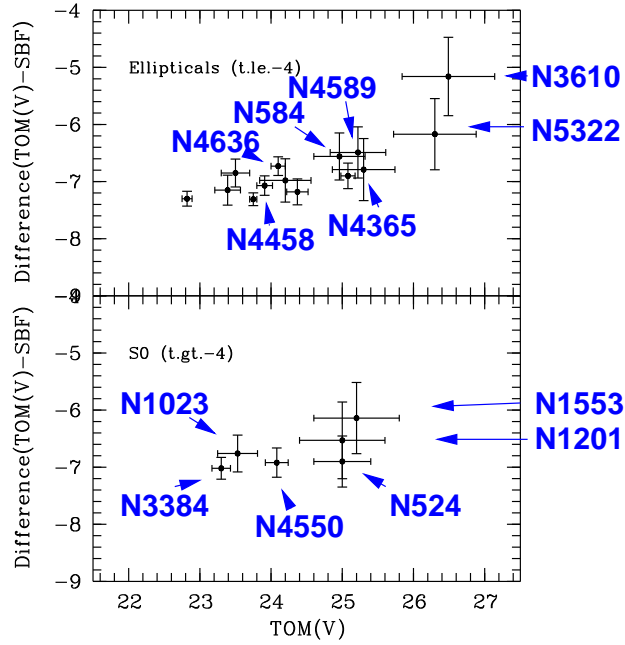}
\end{center}
\caption[]{This graph shows those galaxies, whose GCLF distances deviate more than 
suggested by the measurement uncertainties from SBF distances.  The upper 
panel shows the elliptical galaxies (t-parameter less than -4), the lower
panel the S0-galaxies (t-parameter higher than -4). In the majority of these
objects, one finds evidence that the stellar populations are not exclusively
old. A certain fraction of intermediate-age clusters would make the 
deviation from the SBF distance understandable, as explained in the text.  
  }
\label{deviators}
\end{figure}

Like in human society, deviations from the norm may be sometimes more
interesting and illustrative than reconciliation with it. Let's look at
Fig.\ref{deviators}. Plotted are those TOMs which deviate from the ''universal'' TOM
by more than what is suggested by their uncertainties. Since we see from
Fig. \ref{SBFcomp} that the scatter is by no means symmetric but that the most striking
deviations prefer a TOM, which is systematically fainter than expected from
the SBF distances, only these fainter TOMs are shown.

Among the elliptical galaxies, the largest deviation is shown by NGC 3610, admittedly with a large error. But
more interesting is the fact that this galaxy violates one important condition
for the TOM to be a viable distance indicator, namely that its GC population is
old. NGC 3610 is known to host GCs of intermediate-age. Whitmore et al. 
\cite{whitmore97a}
and Whitmore et al. \cite{whitmore02} estimate an age of about 4-6 Gyr for the red (and
presumably metal-rich)
GCs which plausibly have their origin in a merger event (Schweizer \& Seitzer
\cite{schweizer92}). However, many of the metal-rich
clusters probably had been brought in by the progenitor galaxies. Strader et
al. \cite{strader02} find among their sample of 6 metal-rich clusters only one
with an age of 1-5 Gyr. But since the other clusters are located in the outer
halo, this cannot strongly constrain the fraction of intermediate-age metal-rich clusters.

As we later shall discuss, the GCLF is expected to change its TOM,
resulting in fainter TOMs for younger cluster populations. Indeed, Whitmore et
al. \cite{whitmore02} find a TOM of $25.44\pm0.1$ for the blue clusters alone, while no
TOM is visible at all for the red cluster population. This decreases the difference to the
SBF distance, however, it still remains large. On the other hand,  
the existence of an intermediate-age population also influences the SBF distance
by enhancing  the fluctuation signal (mainly through the enhanced number
of asymptotic giant branch stars) and thus would lead to a spuriously smaller distance without any correction term. Normally, the color V-I is used
to correct for differences in the population (see Liu et al. \cite{liu00} and Blakeslee et al. \cite{blakeslee01} for a
deeper discussion). Whether this correction is always sufficient or fails in
some casesx, cannot be discussed here.  

So the suspicion arises that a difference between the GCLF distance and the SBF
distance may in general be produced by intermediate-age GCs. This conjecture
indeed gets support by looking at other galaxies in  Fig.\ref{deviators}.
 Besides NGC 3610,
younger clusters have been detected in NGC 4365 (Larsen et al. \cite{larsen02}, Puzia et al. \cite{puzia02}, also conjectured to be 
 a merger remnant (Surma \& Bender \cite{surma95}). Note, however, that Davis et al. \cite{davies01} did not find evidence for intermediate-age populations
from their integral-field spectroscopy in the galaxy iself. 

Other ellipticals show strong $H_{\beta}$-lines, indicating as well a younger
population, and NGC 4636 hosted a supernova Ia, whose progenitors should
also be of intermediate age (e.g. Leibundgut \cite{leib00}).

However, there are also examples where the existence of an intermediate-age population is not supported by the present literature. For NGC 3379 and NGC 4472, the difference to the SBF distance is anyway marginal, perhaps still so for NGC 4660. The case of NGC 4291 is
 hard to assess because of the large uncertainties, but NGC 4473 and NGC 5846
pose a problem. The spectroscopic evidence for intermediate-age populations
normally come from the central regions and it may be that the outer parts,
from which the globular clusters are sampled, 
still host younger populations. The other possibility is that either the SBF
distance
or the TOMs are erroneous. In any case one has to wait for further observations.
 
Turning to the S0's, NGC 1023 and NGC 3384  are perhaps candidates 
 for 
hosting intermediate-age populations. Both galaxies
show indications of star formation activity in their inner regions
(Kuntschner et al. \cite{kuntschner01}, Sil'chenko \cite{silchenko99}).
However, the faint TOM of NGC 1023 does not seem to be related to intermediate-age
clusters.  
Larsen \& Brodie \cite{larsen00b}
 identified beside the ''normal'' compact GCs (both red and blue)
a population of faint extended red GCs. The inclusion of these latter objects
in the luminosity function is mainly responsible for the deviating position
of NGC 1023. Leaving them aside results in a distance modulus well agreeing 
with the SBF distance.    
A similar finding is reported for NGC 3384 by Larsen et al. \cite{larsen01}. 
Brodie \& Larsen \cite{brodie02} found that these faint extended clusters belong
to the disk populations of their host galaxies and quote an age of at least
7 Gyr. 

 NGC 4550 contains two
counterrotating stellar disks (Rix et al. \cite{rix92}) and molecular gas has been
detected by Wiklind \& Henkel \cite{wiklind01} which is supposed to have its origin in
a recent accretion event. Similarly striking findings are not reported for  
NGC 1553 or NGC 1201. However, both are shell galaxies (e.g. Longhetti et al.
\cite{longhetti00}), which hints at earlier interactions or mergers.  

NGC 524 again is a supernova Ia host galaxy. 
The question whether the appearance of a supernova of type Ia in an early-type
galaxy always indicates an intermediate-age stellar population is beyond the scope of
this article, but a few remarks on GCSs of Ia host galaxies are appropriate.
NGC 1316 in the Fornax cluster, a merger remnant and host to two SN Ia's, has
a GCS where 2-3 Gyr old clusters have been found, probably formed during the
merger event (Goudfrooij et al. \cite{goud01a}, \cite{goud01b}). Deep VLT and
HST photometry does not reveal a TOM; the GCLF increases
steadily down to beyond the observation limit (Grillmair et al. \cite{grillmair99}, Gilmozzi, this conference). In the work
of G\'omez \& Richtler \cite{gomez01} who quote a TOM which is in good
 agreement with the SBF distance, the TOM was not actually reached, but extrapolated,
and the agreement with the SBF distance perhaps stems from the fact that in the
outer region, where this data has been sampled, the fraction of intermediate-age
clusters is low.

Other early-type Ia host galaxies with investigated GCSs, where 
intermediate-age cluster populations have been identified, are NGC 5018 (SN 2002 dj) (Hilker \&
Kissler-Patig \cite{hilker96}) and NGC 6702
 (SN 2002cs) (Georgakakis et al. \cite{georgakakis01}). Unfortunately,
NGC 6702 is too far
for an analysis of its GCLF and the TOM of the NGC 5018 system must be
largely extrapolated, so it remains uncertain. 

But we also have examples of Ia hosts, where the GCLF distance agrees quite well
or is even smaller than the SBF distance, e.g. NGC 4621 (2001A), NGC 1380 (1992
A), NGC 4526 (1994D) and NGC 3115 (1935B). If there are intermediate-age populations in these galaxies, they do not seem to contaminate the GCLF.  

An interesting note regarding Ia host galaxies and GCSs 
 can be made from the paper of Gebhardt \& Kissler-Patig 
\cite{gebhardt99}. These authors analyze the V-I colour distribution of the GCs
of a sample of early-type galaxies. Their ''skewness'' parameter measures the
asymmetry of the colour distribution with respect to the mean colour. The two
GCSs which are skewed strongest towards red (e.g. metal-rich) clusters both 
belong to Ia host galaxies (NGC 4536, NGC 4374) as well as does the fourth in this
sequence (NGC 3115).

All this, of course, does not mean that in those cases where GCLF and SBF distances agree within the uncertainties,
the stellar populations are necessarily old. However, a comparison with the compilation of galaxy ages by Terlevich \& Forbes \cite{terlevich01} reveals that among the
ellipticals, only NGC 720 (3.4 Gy) is quoted with an age lower than 5 Gyr.  Among the S0's we
 have only NGC 3607 (3.6 Gyr) and NGC 6703 (4.1 Gyr), i.e. strikingly
less candidates for hosting younger populations than among the deviating
ones.
 
Summarizing, it seems that many of the cases where the SBF distance does not
agree with the GCLF distance, can be related to the presence of intermediate-age
populations, particularly among the ellipticals. 
  
Tab.\ref{deviat.tab} lists all galaxies in Fig. \ref{deviators} with their TOM,
its difference with the SBF distance, references for the TOM and a reference
for other properties of the host galaxy.

\begin{table}
\caption{A list of galaxies whose TOMs indicated larger distances than the
SBF distances beyond the
uncertainty limits. For many of these galaxies one finds evidence for the
existence of intermediate-age (IM) populations.  
}

\begin{center}
\renewcommand{\arraystretch}{0.9}
\setlength\tabcolsep{8pt}
\begin{tabular}{lllll}
\hline\noalign{\smallskip}

     Name &  TOM &   diff(TOM-SBF) &  Ref. &   Remarks \\  
\hline
  
 & Ellipticals & & & \\
\hline
  N3610   & $26.49 \pm 0.65 $ & $-5.16 \pm 0.69 $& \cite{kundu01},\cite{whitmore02} &  IM clusters\\ 
  N5322   & $26.30 \pm 0.58 $ & $-6.17 \pm 0.62 $& \cite{kundu01},\cite{proctor02} &  IM age\\
  N4589   & $25.22 \pm 0.39 $ & $-6.49 \pm 0.45 $& \cite{kundu01},\cite{trager99}  &  $H\beta$ strong\\
  N0584   & $24.96 \pm 0.36 $ & $-6.56 \pm 0.41 $& \cite{kundu01},\cite{kodama98},\cite{trager99}&$H\beta$ strong\\
  N4636   & $24.10 \pm 0.10 $ & $-6.73 \pm 0.16 $& \cite{kissler94}   &  Ia host\\
  N4291   & $25.30 \pm 0.44 $ & $-6.79 \pm 0.54 $& \cite{kundu01}     &  old?\\
  N4697   & $23.50 \pm 0.20 $ & $-6.85 \pm 0.24 $& \cite{kavelaars00},\cite{proctor02} &  IM age\\
  N5846   & $25.08 \pm 0.10 $ & $-6.90 \pm 0.22 $& \cite{forbes96},\cite{kuntschner01}  &  old?\\
  N4458   & $24.20 \pm 0.36 $ & $-6.98 \pm 0.38 $& \cite{kundu01},\cite{kuntschner01}  &  $H\beta$ strong\\
  N4473   & $23.91 \pm 0.11 $ & $-7.07 \pm 0.17 $& \cite{kundu01},\cite{kuntschner01}  &  old?\\
  N4660   & $23.39 \pm 0.18 $ & $-7.15 \pm 0.26 $& \cite{kundu01},\cite{kuntschner01}  &  old?\\
  N4365   & $24.37 \pm 0.15 $ & $-7.18 \pm 0.23 $& \cite{larsen01},\cite{puzia02},\cite{surma95} &   IM clusters\\
  N3379   & $22.82 \pm 0.07 $ & $-7.30 \pm 0.13 $& \cite{kundu01},\cite{kuntschner01} &  old?\\
  N4472   & $23.75 \pm 0.05 $ & $-7.31 \pm 0.11 $& \cite{kundu01},\cite{kuntschner01}  &  old?\\
\hline
& S0  & & & \\
\hline
  N3384  & $23.30 \pm 0.13$ &$ -7.02 \pm 0.19$ & \cite{larsen01},\cite{kuntschner01} & extended GCs, $H\beta$ strong \\
  N4550  & $24.08 \pm 0.16$ &$ -6.92 \pm 0.26$ & \cite{kundu01},\cite{wiklind01} &     molecular gas, merger\\
  N0524  & $25.00 \pm 0.40$ &$ -6.90 \pm 0.45$ & \cite{kundu01b}       &   Ia host\\
  N1023  & $23.53 \pm 0.28$ &$ -6.76 \pm 0.32$ & \cite{larsen00b},\cite{brodie02},\cite{silchenko99}   & extended GCs, IM nucleus\\
  N1201  & $25.00 \pm 0.60$ &$ -6.53 \pm 0.67$ & \cite{longhetti00}  &  shell galaxy\\
  N1553  & $25.20 \pm 0.60$ &$ -6.14 \pm 0.62$ & \cite{longhetti00}  &  shell galaxy\\
\hline

\end{tabular}
\end{center}

\label{deviat.tab}
\end{table}

\section{Why does it work?}

\begin{figure}
\begin{center}
\includegraphics[width=.9\textwidth,angle=-00]{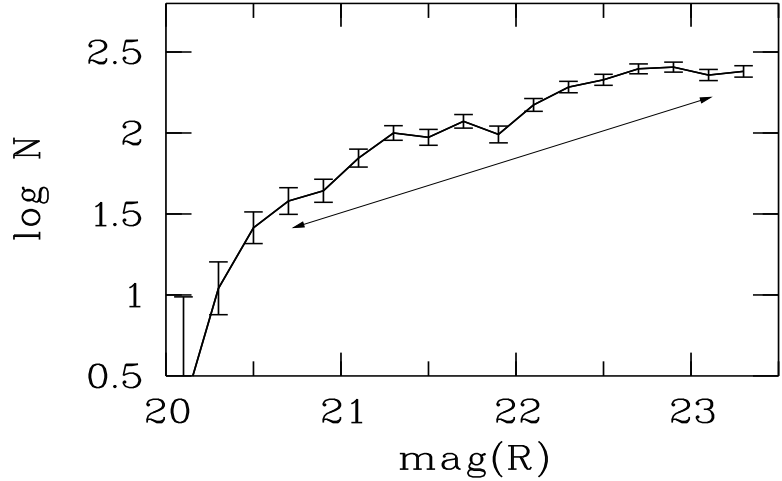}
\end{center}
\caption[]{This plot shows the luminosity function in the R-band for the GCS of NGC 1399, the
central galaxy in the Fornax cluster (Dirsch et al. \cite{dirsch03}). This luminosity function
comprises about 2600 clusters brighter than R=23 and thus is one
of the best available. For magnitudes fainter than R=20.5, the linear
luminosity function is well represented by a power-law with an exponent of
about -2 (the slope s in the present diagram is related to the power-law
exponent $\alpha$ by $\alpha$ = s/0.4 - 1). It steepens considerably for brighter magnitudes. However, a representation
by a log-normal function for the entire magnitude range is as good.
}
\label{LF1399}
\end{figure}

What could be the reason for an universal TOM of old GCSs? A globular cluster with $M_V = -7.5$ mag 
has a mass of about 150000 $M_{\odot}$, adopting an average $M/L_V$ of 2.5
 (Pryor \& Meylan \cite{pryor93}).  Is this particular mass somehow distinguished?
 One has to realize that the
magnitude scale is logarithmic. Binning in linear luminosity units  instead of magnitudes,
we would not see any striking feature at the luminosity corresponding to
the TOM.  After remarks by Surdin \cite{surdin79}, Racine \cite{racine80}, and Richtler \cite{richtler93},
 regarding the power-law
nature of the linear luminosity function of galactic GCs, McLaughlin \cite{mclaughlin94} put this concept on a formal basis.  If the  luminosity function
can be described as $NdL \sim L^{\alpha(L)}$, where N is the number of clusters
found in the luminosity interval L+dL and $\alpha(L)$ a function
of  L, then one has in magnitudes $N dM_V \sim 10^{0.4 M_V (1-\alpha)}$,
i.e. 
 the TOM is found where $\alpha(L)$ just changes from smaller than -1 to larger
than -1.  So the location of the TOM
does not express a
 specific physical property at this particular mass. However, the underlying universal property must be a universal mass function. Harris \& Pudritz \cite{harris94} first investigated the mass function of GCSs of different galaxies,
assuming a constant M/L.
They found that such diverse
systems as that of the Milky Way and of M87 can be described  
 by a common power-law exponent of $\alpha \approx -1.8$ for masses 
higher than about $10^5$ solar masses. Larsen et al. \cite{larsen01} 
found in their larger sample on the average $\alpha = -1.74 \pm 0.04$ between $10^5$ and $10^6$
solar masses. However, in very rich
GCSs, such as that of M87 or NGC 1399, the slope becomes distinctly steeper for cluster
masses larger than about $10^6$ solar masses (see Fig.\ref{LF1399}).

Ten years ago, GCSs had been almost exclusively associated with old stellar
populations. Meanwhile, systems of young globular clusters have been detected
in many merging galaxies, the most prominent ones being the Antennae NGC  
4038/4039 (Whitmore \& Schweizer \cite{whitmore95} ) and NGC 7252 (Whitmore et al. 1993)
 (see Whitmore \cite{whitmore00} for a complete listing until 2000), but also in normal
spiral galaxies (Larsen \& Richtler \cite{larsen99}, Larsen \& Richtler \cite{larsen00a}).

Determinations of the luminosity functions resulted so far consistently in
power-laws with an exponent of about -2, without compelling indications that this
exponent changes over the observed luminosity range as in the case of the
GCSs of giant ellipticals (Whitmore \cite{whitmore00}), given the uncertainties caused
by internal extinction and by the age spread among a cluster system.
There was some debate regarding the mass function of GCs in the the Antennae as
derived from the luminosity function. 
The Antennae may show a bend at about $M_V =
-11$, becoming steeper towards the bright end (Whitmore \cite{whitmore00}, Zhang \& Fall \cite{zhang99}).
Fritze-v. Alvensleben \cite{fritze99}) found a log-normal mass distribution like
for old systems, which was contradicted by Zhang and Fall (\cite{zhang99}), who attributed this
difference to the effect of varying extinction and ages, and found a uniform
power-law. 
 If we assume that young GCs
are born obeying  a universal luminosity function like $dN/dL \sim L^{-2}$,
and accordingly with a mass function of the same shape, then we must ask,
what processes can transform such a luminosity function into the approximately
log-normal luminosity functions of old GCSs. If these processes work in a
universal manner, then the universality of the TOM could be explained.

\section{How does an initial cluster mass function change with time?}


\begin{figure}
\begin{center}
\includegraphics[width=.9\textwidth]{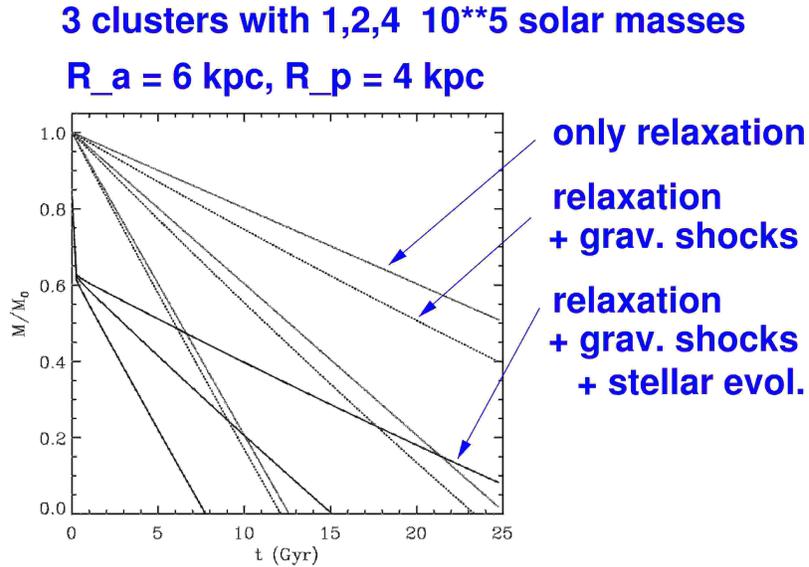}
\end{center}
\caption[]{This plot is taken from Fall \& Zhang \cite{fall01}. It shows for three different
clusters ($10^5, 2\cdot 10^5, 4\cdot 10^5$ solar masses) the cumulative growth of the relative
mass loss by two-body relaxation
(upper line), two-body relaxation plus tidal shocks (middle line), and added to that the mass
loss from stellar evolution (lower line).
}
\label{fall.ps}
\end{figure}

A young star cluster is exposed to different destruction mechanisms. If it
is still young, mass loss from massive star evolution plays an important role.
At later times, two-body relaxation, dynamical friction, and tidal shocks,
when the cluster enters the bulge region of its host galaxy or moves through
a disk, can
be efficient in decreasing the cluster's mass, depending on its mass, it's density
and it's orbit in the host galaxy. The most general statement is that low-mass
clusters are more affected by disruption processes than high-mass clusters, so
an initial power-law of the mass distribution is more strongly distroyed on the low
mass end and may develop a shape which finally resembles a log-normal distribution.  

\begin{figure}
\begin{center}
\includegraphics[width=.8\textwidth]{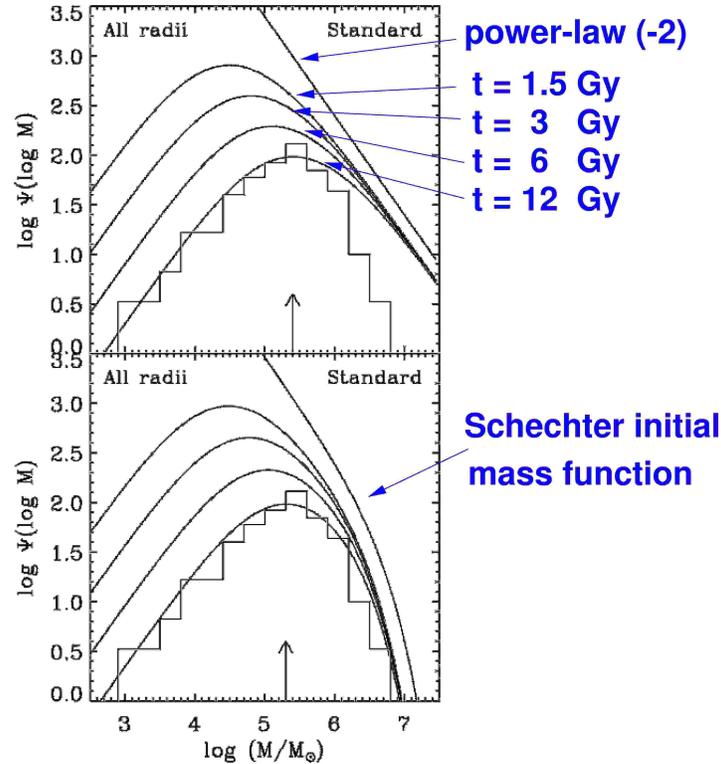}
\end{center}
\caption[]{This plot is taken from Fall \& Zhang \cite{fall01}. It shows the evolution
of a cluster system for two different initial mass functions: a power-law with an exponent
of -2 (upper panel) and a Schechter function (lower panel).
}
\label{fallevolution}
\end{figure}

Many people have worked on this problem, among them Aguilar et al. \cite{aguilar88},
 Okazaki \& Tosa \cite{oka95}, Elmegreen \& Efremov \cite{elm97}, Gnedin \& Ostriker \cite{gnedin97}, Murali \& Weinberg (1997a,b,c) Vesperini \cite{vespe98},\cite{vespe00}, 
 Fall \& Zhang \cite{fall01}. We cannot present all work in detail, instead we
choose the analytical model by Fall \& Zhang in order to illustrate the most important
results. Fig. \ref{fall.ps} (Fig. 1 of Fall \& Zhang 2001) shows the time evolution  for three different masses for a cluster which is on a slightly elongated orbit. The dotted lines indicate the effect of two-body relaxation alone. The dashed lines
additionally include gravitational shocks, and the solid lines add the effect
of mass loss by stellar evolution.

Under a wide variety of conditions, the mass function of a GCS develops a
peak which is progressively shifted to higher masses, as the evolution of
the cluster system proceeds. After, say, 12 Gyr, Fall \& Zhang get from
their  model a peak mass (in logarithmic bins) which may well represent the mass corresponding to the TOM observed in the Milky Way or in elliptical galaxies (Fig.\ref{fallevolution}).

 However,
it seems that the assumption of a power-law with an exponent around -2, as
suggested by the young cluster systems in merging galaxies, cannot reproduce
well the log-normal shape in the mass-rich domain observed in many galaxies.
This is because the shape of the mass function above a few times $10^6 M_{\odot}$ practically does not change by evolutionary processes. 
Instead, an initial log-normal mass function  works much better in resembling
  the bright end of the luminosity function of ellipticals (Vesperini \cite{vespe00},
\cite{vespe01}) (but see the section on the brightest clusters).


The dynamical evolution of a GCS may raise doubts on the general quality of
the GCLF as a distance indicator, if the evolutionary history of a
GCS is not negligible. The GCLF might also depend on whether the TOM is measured
at small or large galactocentric radii. In the inner parts of a galaxy, the
TOM is expected to be brighter. Gnedin \cite{gnedin97} finds significant differences
in this sense for the Milky Way, M31, M87, which for M31 has been confirmed
by the improved sample of Barmby et al. \cite{barmby01}.
Also the brighter TOM (with respect to the SBF distance) of
the Sombrero may have its explanation in the dynamical history of this GCS.
The Sombrero possesses an extraordinary large bulge, where dynamical shocks might work
more efficient than in other galaxies (naively assuming, of course, that the
SBF distance is correct). Note, however, that the HST-observations by Larsen et al. 
\cite{larsen00b} reveal a GCLF for the Sombrero whose TOM is not very well defined.

\begin{figure}
\begin{center}
\includegraphics[width=.6\textwidth,  angle=0]{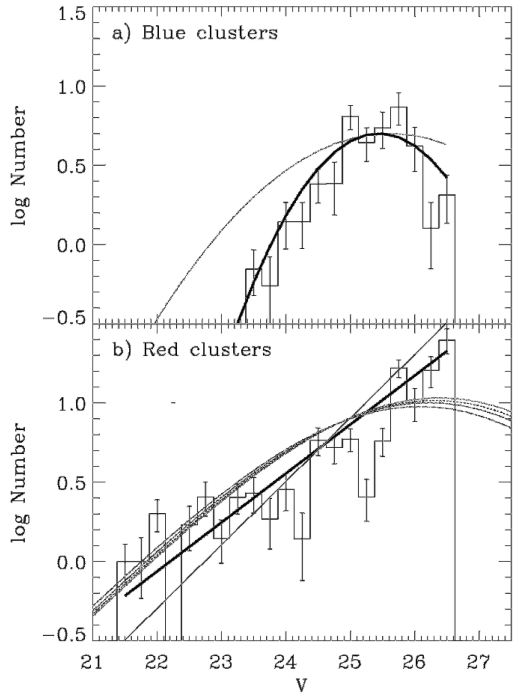}
\end{center}
\caption[]{This plot is taken from Whitmore et al. \cite{whitmore02}. It shows
the luminosity functions (LFs) of blue (upper panel) and red (lower panel) globular
 clusters in NGC 3610. 
The blue clusters exhibit a TOM at V = 24.55. The solid line 
marks a Gaussian fit with $\sigma = 0.66$, the dotted line with $\sigma = 1.1$,
which obviously do not fit. The LF of the red clusters increases
with no sign of a flattening. The thick solid line is a power-law fit with
$\alpha = -1.78$. The other curves are model LFs based on
Fall \& Zhang \cite{fall01} in combination with population synthesis models
of Bruzual \& Charlot (unpublished). The thin solid line is the zero-age LF
(power-law with $\alpha = -2$),  
 the others correspond to ages of 1.5 Gyr, 3 Gyr, 6 Gyr, and 12 Gyr (from top to bottom).
The TOMs are hardly distinguishable because in these models the brightening of the TOM by dynamical
evolution is balanced by the fading due to stellar evolution.}
\label{N3610}
\end{figure}

One of the best investigated galaxies among those which shows a marked
 difference
between the GCLF and the SBF distance is the elliptical galaxy NGC 3610.
  Scorza
\& Bender \cite{scorza90} found a disk and other morphological signatures indicating previous interaction or a merger event.  
It's location in Fig.\ref{SBFcomp} corresponds to the TOM quoted by Kundu \& Whitmore \cite{kundu01}.
In a subsequent paper, Whitmore et al. \cite{whitmore02} performed a more detailed investigation of the
GCS of NGC 3610, based on new HST data. Fig. \ref{N3610} shows the LFs separately for the
blue and the red clusters. While for the blue clusters the TOM is measured to be at $V = 25.44
\pm 0.1$, the red clusters show a LF rising until the photometry limit. Whitmore et al. combined
the destruction model of Fall \& Zhang with evolutionary models of stellar populations. The
resulting model LFs are indicated in the lower panel. The data are not yet deep enough to
show a turn-over for the red clusters, which by the models is predicted to be at around
$V \sim 26 $.  Whitmore et al. state that in the context of the Fall \& Zhang models, the
brightening of the TOM during the dynamical evolution of the cluster system is almost completely balanced by the fading of the stellar population during this time.  This may well be
an explanation for the universality of the TOM.
However, given the approximate nature of the analytic models of Fall \& Zhang and the
the dependence of the destruction processes on the actual environment, this probably
does  not apply to every galaxy.

We therefore can conclude that in the case of NGC 3610, a large part of the deviation of
the GCLF distance from the SBF distance comes from the fact that the contribution of the
presumably younger red clusters causes a fainter TOM than from the blue, metal-poor and
presumably older clusters alone. But also when we use only the TOM of the blue clusters to
determine the distance modulus, which then would be 32.94, a significant difference remains
to the SBF modulus, which is 31.65. This cannot be resolved here. The modelling of SBF's
accounts for the population structure (Liu et al. \cite{liu02}, Blakeslee et al. 
\cite{blakeslee01}) but may fail in extreme cases.

\section{The brightest clusters}
 
Regarding the significance of the GCLF as a distance indicator, its shape
in the domain of the  
brightest clusters is less important. But since  
dynamical models indicate that the GCLF for clusters more 
massive than about $10^6 M_{\odot}$ is not modified by destruction processes,
they bear potential information about the formation of a GCS. Two different
views on a GCLF like that of Fig.\ref{LF1399} exist: It can be seen as a power-law with an exponent around -2 with a cut-off at higher masses or it
can be seen as a log-normal function.

 Adopting the first view,
the GCLF would resemble in large parts the LF found in young cluster systems. The cut-off
at high masses may have different reasons. Possibly these very massive clusters (the
brightest clusters in NGC 1399 have about $10^7 M_{\odot}$) are not globular
clusters in the normal sense, but the dynamically stripped nuclei of dwarf
galaxies. In the Milky Way, we have $\omega$ Centauri as a possible example (e.g.
Hilker \& Richtler \cite{hilker00}). In this case, the LF at the bright end
would depend on the accretion rate of dwarf galaxies, which pausibly is 
highest in massive galaxies in dense environments like NGC 1399 or M87.
Or the formation history of the most massive clusters is principally different
from the less massive ones. The peculiar cluster in NGC 6946 (Larsen et
al. \cite{larsen01a}, \cite{larsen02a}) with a mass of about $10^6 M_{\odot}$
and an age of 15 Myr is surrounded by a round, star forming complex
 of about 600 pc diameter, which gives the impression of a disk-like 
structure with  the massive cluster near its center. Such a configuration 
suggests that 
the cluster mass is determined or partly determined by accretion
from a larger region, resulting in a  steeper mass function for
massive clusters.
 
Taking the second view of a log-normal function, one has the possibility to
relate such a shape to coagulation processes by which GCs
might have been formed through the merging of smaller subunits. Based on
ideas by Harris \& Pudritz \cite{harris94}, McLaughlin \& Pudritz \cite{mclaughlin96} developed a model, in which GCs form inside the cores of supergiant
 molecular clouds. These cores are built up by internal collisions and
subsequent coagulation of smaller 
clouds. Star formation tends to partly disrupt these cores and in an equilibrium
between coagulation and disruption, a mass spectrum of cores results, which
directly resembles the GC mass spectrum. This is because the formation of 
GCs in these cores must occur with a high star formation efficiency in order
for the GC to stay bound, i.e. the mass of the core is closely related to the mass
of the final cluster. 

See also Burkert \& Smith \cite{burkert00} who argue that the mass spectra of
GCSs can be fitted with a form $dN/dm \sim m^{-3/2} \cdot exp(-m/m_c)$, where
$m_c$ is a ''truncation mass``. Such a shape resembles the long-time solution
of the coagulation model of Silk \& Takahashi \cite{silk79}, initially starting
with small progenitor clouds of equal mass.

Although the relation between the GC mass spectrum  and the mass spectrum
of the progenitor clouds is open to speculation, the power-law interpretation
of the GCLF has some attractive features over the log-normal law interpretation.It relates the GCLF of old clusters system with young ones, and it offers
a simple explanation by a direct link to the mass spectrum of molecular
clouds. 
 
It is amazing that molecular clouds in the Galaxy exhibit a mass spectrum resembling so
closely that of GCs. See the introductory part of Elmegreen \cite{elm02}
for a compilation of references. This power-law behaviour may be the result
of a fractal structure of the interstellar gas  caused by turbulence and
selfgravitation   
(Fleck \cite{fleck96}, Elmegreen \& Falgarone \cite{elm97}, Elmegreen \cite{elm02}). Therefore the universality of the GCLF probably has its ultimate explanation
in the universality of the interstellar gas structure.
  

 
\section{Conclusions}

We have seen that the method of globular cluster luminosity functions (GCLFs) allows one
 to determine distances to
early-type galaxies, which are as accurate as those derived from
surface brightness fluctuations (SBFs), once the conditions of high data quality
and sufficiently rich cluster systems are fullfilled. The achievable accuracy
of distance moduli is of the order 0.2 mag. The absolute turn-over
magnitudes (TOMs), if calibrated by SBF distances, agree very well with
those of the globular cluster systems of the Milky Way and the Andromeda nebula.
Therefore the TOM is indeed a universal property of old globular cluster systems. 

The comparison of SBF distances with GCLF distances reveals however many
discrepant cases, in which the GCLF distances are systematically larger than
the SBF distances beyond the limits given by the uncertainties. In some elliptical
 galaxies, direct evidence for the existence of intermediate-age globular 
clusters is available. In others, intermediate-age stellar populations are
indicated by a variety of findings, which again may suggest a certain 
fraction of intermediate-age globular clusters as well. The S0-galaxies NGC
1023 and NGC 3384 exhibit a population of faint extended red clusters, which
cause a fainter TOM, if they are included in the luminosity function. 

That a globular cluster system, consisting mainly of old clusters, in which intermediate-age clusters are mixed in, exhibits a fainter TOM, can be understood 
by the dynamical evolution of cluster systems. Young globular clusters, which
are found in large numbers  in merging galaxies, 
are formed 
according to  a power-law mass function with an exponent around -2. The cluster
system then undergoes a dynamical evolution where the mass loss of 
individual clusters is caused  by two-body relaxation, tidal shocks and mass loss by stellar evolution. This results in a preferential destruction of
low-mass clusters, modifying the initial power-law mass function in such a
way that the corresponding luminosity function on the magnitude scale shows a TOM which
becomes brighter as the system evolves. The fading of clusters by stellar
evolution counteracts this brightening to some degree.  

The universality of the GCLF probably has its origin in the fractal structure of the
interstellar medium, which results in a power-law mass spectrum for molecular clouds
with an exponent of -2, similar to that found for young globular cluster systems.

\section{Acknowledgements}

Thanks go to Doug Geisler for a careful reading of the manuscript and helpful
criticism. I am indebted to S{\o}ren Larsen for indicating unclear points and
calling attention to NGC 1023 and NGC 3384.
Many discussions with Boris Dirsch are gratefully acknowledged as well as
 support by the FONDAP Center for Astrophysics, Conicyt 15010003.

\end{document}